\documentclass[11pt,reqno]{amsart}
\usepackage[top=1in, left=1in, right=1in, bottom=1in]{geometry}
\usepackage[parfill]{parskip}    
\usepackage{algorithm}
\usepackage{algpseudocode}
\usepackage{hyperref}
\usepackage{url}
\usepackage{verbatim}
\usepackage{amssymb}
\usepackage{amsmath}
\usepackage{amsaddr}
\usepackage{enumitem}
\usepackage{setspace}
\usepackage{natbib}
\usepackage{algpseudocode}
\usepackage{graphicx}
\usepackage{times}

\algnewcommand{\Inputs}[1]{%
  \State \textbf{Inputs:}
  \Statex \hspace*{\algorithmicindent}\parbox[t]{.8\linewidth}{\raggedright #1}
}
\algnewcommand{\Initialize}[1]{%
  \State \textbf{Initialize:}
  \Statex \hspace*{\algorithmicindent}\parbox[t]{.8\linewidth}{\raggedright #1}
}

\title[VI for RVD]{Variational inference for rare variant detection in deep, heterogeneous next-generation sequencing data}
\author[F. Zhang AND P. Flaherty]{Fan Zhang\,$^{1}$, Patrick Flaherty\,$^{1,2}$}
\address{$^{1}$Department of Biomedical Engineering, Worcester Polytechnic Institute, MA, USA\\
$^{2}$Department of Mathematics and Statistics, University of Massachusetts, Amherst, MA, USA}

\begin{document}

\maketitle

\begin{abstract}
  The detection of rare variants is important for understanding the genetic heterogeneity in mixed samples.
Recently, next-generation sequencing (NGS) technologies have enabled the identification of single nucleotide variants (SNVs) in mixed samples with high resolution.
Yet, the noise inherent in the biological processes involved in next-generation sequencing necessitates the use of statistical methods to identify true rare variants.
We propose a novel Bayesian statistical model and a variational expectation-maximization (EM) algorithm to estimate non-reference allele frequency (NRAF) and identify SNVs in heterogeneous cell populations.
We demonstrate that our variational EM algorithm has comparable sensitivity and specificity compared with a Markov Chain Monte Carlo (MCMC) sampling inference algorithm, and is more computationally efficient on tests of low coverage ($27\times$ and $298\times$) data.
Furthermore, we show that our model with a variational EM inference algorithm has higher specificity than many state-of-the-art algorithms.
In an analysis of a directed evolution longitudinal yeast data set, we are able to identify a time-series trend in non-reference allele frequency and detect novel variants that have not yet been reported.
Our model also detects the emergence of a beneficial variant earlier than was previously shown, and a pair of concomitant variants.
\end{abstract}

\section{Introduction}
Massively parallel sequencing data generated by next-generation sequencing technologies is routinely used to interrogate SNVs in research samples ~\citep{koboldt2013next}.
For example, deep sequencing confirmed the degree of genetic heterogeneity of HIV and influenza ~\citep{flaherty2011ultrasensitive, ghedin2011deep}.
Many solid tumors are represented as having intra-tumor heterogeneity by DNA sequencing technology, where SNVs are rare in the population ~\citep{navin2010inferring}.
Also, whole genome sequencing has revealed that many beneficial mutations of minor allele frequencies are essential to respond to dynamic environments ~\citep{kvitek2013whole}.
However, rare SNVs identification in heterogeneous cell populations is challenging, because of the intrinsic error rate of next generation sequencing platform ~\citep{shendure2008next}.
Thus, there is a need for accurate and scalable statistical methods to uncover SNVs in heterogeneous samples.

A number of computational methods have been developed to detect SNVs in large scale genomic data sets.
These methods can be roughly categorized as probabilistic or heuristic or some combination.
Among all of the current probabilistic methods, the Bayesian probabilistic framework has been increasingly used to calculate the unobserved quantities such as variant allele frequency given the observed genomic sequencing data.
GATK ~\citep{mckenna2010genome} and SAMTools ~\citep{li2009sequence} use a naive Bayesian decision rule to call variants.
EBCall models sequencing errors based on a Beta-Binomial distribution, where the parameters and latent variables are estimated from a set of non-paired normal sequencing samples ~\citep{shiraishi2013empirical}.
However, the error rate of normal sequencing samples could be unmatched with the error rate of the target samples, which may cause a problem of making false negatives calls ~\citep{wang2013detecting}.
CRISP compares aligned reads across multiple pools to obtain sequencing errors, and then distinguishes true rare variants from the sequencing errors ~\citep{bansal2010statistical}.
However, the bottleneck of CRISP is its low computational efficiency due to a calculation of a large amount of contingency tables.

Generally, independent modeling method models the genotypes of tumour and normal samples separately and then looks for the difference between them.
In contrast to independently analyzing a tumour-normal pair, joint modeling method models a joint-genotype of the tumour and normal samples simultaneously.
JointSNVMix introduces two Bayesian probabilistic models (JointSNVMix1 and JointSNVMix2) to jointly analyze a tumour-normal paired allelic count of NGS data ~\citep{roth2012jointsnvmix}.
JointSNVMix derives an expectation maximization (EM) algorithm to calculate maximum a-posteriori (MAP) estimate of latent variables in a particular probabilistic graphical model.
Furthermore, \citet{roth2012jointsnvmix} reveals that the joint modeling method, JointSNVMix1, observes 80-fold reduction of false positives compared with its independent analogue (SNVMix1).
SomaticSniper ~\citep{larson2012somaticsniper} models the joint diploid genotype likelihoods for both tumour and normal samples.
Additionally, Strelka ~\citep{saunders2012strelka} models the joint probabilistic distribution of allele frequencies for both tumour and normal samples, which is demonstrated to be more accurate compared with the methods based on the estimated allele frequency tests between tumour and normal samples.

An alternative strategy to probabilistic methods is heuristic methods that use a set of criteria to select variant positions instead of modeling the data using probabilistic distributions.
VarScan is an example of a heuristic method that compares tumour and normal samples by satisfying some lower bounds, such as a certain variant allele frequency and a number of allele counts ~\citep{koboldt2012varscan}.
SNVer focuses on a frequentist method that is able to calculate $P$-values, but \citet{wei2011snver} pointed out that this approach fails to model sampling bias that will reduce the power of detecting true rare variants.

In previous work, we developed a Beta-Binomial model to characterize a null hypothesis error rate distribution at each position.
Using this rare variant detection (RVD) model, we call rare variants by comparing the error rate of the sample sequence data to a null distribution obtained from sequencing a known reference sample ~\citep{flaherty2011ultrasensitive}.
RVD can identify mutant positions at a 0.1\% fraction in mixed samples using high read depth data.

We improved upon that work, RVD2, by using hierarchical priors to tie parameters across positions~\citep{he2015rvd2}.
We derived a Markov Chain Monte Carlo (MCMC) sampling algorithm for posterior inference.
However, the main limitation of MCMC is that it is hard to diagnose convergence and may be slow to converge~\citep{jordan1999introduction}.
An alternative method, that we explore here, is to use variational inference, which is based on a proposed variational distribution over latent variables.
By optimizing variational parameters, we fit an approximate distribution that is close to the true posterior distribution in the sense of the Kullback-Liebler (KL) divergence.
Variational inference can now handle nonconjugate distributions and tends to be more computationally efficient than MCMC sampling~\citep{peterson1989explorations}.

Here, we propose a variational EM algorithm for our Bayesian statistical model to detect SNVs in heterogeneous NGS data.
We show that variational EM algorithm has comparable accuracy and efficiency compared with MCMC in a synthetic data set.
In section 2, we define the model structure.
In section 3, we derive our variational EM algorithm to approximate the posterior distribution over latent variables.
In section 4, we call a variant by a posterior difference hypothesis test between the key model parameters of a pair of samples.
In section 5, we compare the performance of the variational EM inference algorithm to the MCMC sampling method and the state-of-the-art methods using a synthetic data set.
We also show that our variational EM algorithm is able to detect rare variants and estimate non-reference allele frequency (NRAF) in a longitudinal directed evolution experimental data set.

\section{Model Structure}
\begin{figure}[htpb]
\centering
\includegraphics[width=0.5\textwidth]{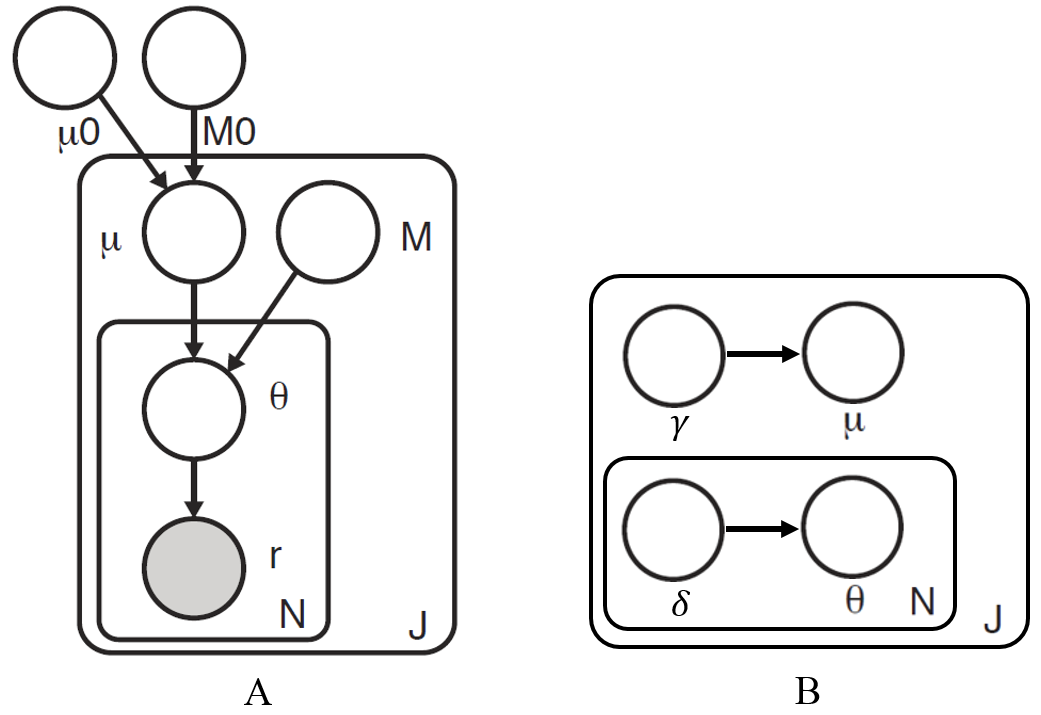}
\caption{A. Graphical model representation of the model.
B. Graphical model representation of the variational approximation  to approximate the posterior distribution.
Observed random variables are shown as shaded nodes and latent random variables are unshaded.
The object of inference for the variational EM algorithm is the joint distribution $p(\mu, \theta|r, n)$.}
\label{tbl:graphical_model}
\end{figure}

Our Bayesian statistical model is shown as a graphical model in Figure~\ref{tbl:graphical_model}.
In the model, $r_{ji}$ is the number of reads with a non-reference base at location $j$ in experimental replicate $i$; $n_{ji}$ is the total number of reads at location $j$ in experimental replicate $i$.
The model parameters are:
\begin{description}[noitemsep]
  \item[$\mu_0$] a global non-reference read rate that captures the error rate across all the positions,
  \item[$M_0$] a global precision that captures the variation of the error rate across positions in a sequence,
  \item[$M_j$] a local precision that captures the variation of the error rate at position $j$ across different replicates.
\end{description}
The latent variables are:
\begin{description}[noitemsep]
  \item[$\mu_j \sim \text{Beta}(\mu_0, M_0)$] position-specific non-reference read rate for position $j$,
  \item[$\theta_{ji} \sim \text{Beta}(\mu_j, M_j)$] the non-reference read rate for position $j$ in replicate $i$.
\end{description}

In Figure~\ref{tbl:graphical_model}B, $\gamma$ is the parameter for the variational distribution for latent variable $\mu$,
and $\delta$ is the parameter for the variational distribution for latent variable $\theta$.
We describe $q(\mu)$ and $q(\theta)$ in detail in section 3.3

The model generative process is as follows:
\begin{enumerate}[noitemsep]
    \item For each location $j \in [1, \ldots, J]$:
	\begin{enumerate}
		\item Draw an error rate $\mu_j \thicksim \text{Beta}(\mu_0, M_0)$
		\item For each replicate $i \in [1, \ldots, N]$:
		\begin{enumerate}
			\item Draw $\theta_{ji} \thicksim \text{Beta}(\mu_j, M_j)$
			\item Draw $r_{ji} | n_{ji} \thicksim \text{Binomial}(\theta_{ji}, n_{ji})$
		\end{enumerate}
	\end{enumerate}
\end{enumerate}

The joint distribution $p(r, \mu, \theta|n; \phi)$ given the parameters can be factorized as
\begin{equation}
    p(r, \mu, \theta| n; \phi) = p(r |\theta, n)p(\theta |\mu; M )p(\mu;\mu_0, M_0).
\end{equation}

\section{Variational Expectation Maximization (EM) Inference}
We developed a non-conjugate variational inference algorithm to approximate the posterior distribution,
\begin{equation}
	p(\mu, \theta | r, n; \phi)  = \frac{ p(r, \mu, \theta| n; \phi) } {p ( r | n; \phi)},
\end{equation}
where the parameters are $\phi \triangleq \{\mu_0, M_0, M\}$.
\subsection{Factorization}
We propose the following factorized variational distribution to approximate the true posterior over latent variables $\mu_j$ and $\theta_{ji}$.
Here, $q(\mu_j)$ approximates the variational posterior distribution of $\mu_j$, which represents the local error rate distribution in position $j$ across different replicates;
and $q(\theta_{ji})$ approximates the posterior distribution of $\theta_{ji}$, which is the error rate distribution in position $j$ replicate $i$.
\begin{equation}
  q(\mu, \theta) = q(\mu)q(\theta) = \prod_{j=1}^J q(\mu_{j}) \prod_{i=1}^N q(\theta_{ji}).
  \label{eq:vardist}
\end{equation}
\subsection{Evidence Lower Bound (ELBO)}
Given the variational distribution, $q$, the log-likelihood of the data is lower-bounded according to Jensen's inequality,
\begin{equation}
\begin{split}
\log p \left( r | \phi \right) &= \log \int_\mu \int_\theta p\left(r,\mu,\theta |n; \phi \right) d\theta d\mu \\
&= \log \int_\mu \int_\theta p\left(r,\mu,\theta |n; \phi \right)\frac{q\left(\mu,\theta \right) }{q\left(\mu,\theta \right) } d\theta d\mu \\
&\geq \int_\mu \int_\theta q\left(\mu,\theta \right) \log \frac{ p\left(r,\mu,\theta |n; \phi \right)}{q\left(\mu,\theta \right)} d\theta d\mu \\
&= E_q \left[ \log p\left(r,\mu,\theta |n; \phi \right)\right] - E_q \left[ \log q\left(\mu,\theta \right)\right] \\
&\triangleq \mathcal{L}(q, \phi).
\end{split}
\end{equation}

The function $\mathcal{L}(q, \phi)$ is the evidence of lower bound (ELBO) of the log-likelihood of the data, which is the sum of $q$-expected complete log-likelihood and the entropy of the variational distribution $q$.
The goal of variational inference is to maximize the ELBO.
Equivalently, $q$ is chosen by minimizing the KL divergence between the variational distribution and the true posterior distribution.

The ELBO can be expanded as
\begin{equation}
\begin{split}
\label{L}
\mathcal{L}(q, \phi) &= E_q \left[ \log p\left(r,\mu,\theta | n; \phi \right)\right] - E_q \left[ \log q\left(\mu,\theta \right)\right] \\
&= E_q \left[ \log p\left(r | \theta, n \right)\right] + E_q \left[ \log p\left(\theta | \mu; M \right)\right] + E_q \left[ \log p\left(\mu ; \mu_0, M_0 \right)\right]- E_q \left[ \log q\left(\mu \right)\right]- E_q \left[ \log q\left(\theta \right)\right]. \\
\end{split}
\end{equation}
We write out each component below.
The detail of the derivation of $E_q \left[ \log p\left(r | \theta, n \right)\right]$, $E_q \left[ \log p\left(\mu ; \mu_0, M_0 \right)\right]$, and $E_q \left[ \log p\left(\theta | \mu; M \right)\right]$ is shown in Appendix~\ref{appendix:ELBO}.
\begin{equation}
\begin{split}
\label{r}
E_q \left[ \log p\left(r | \theta, n \right)\right] &= \sum_{j=1}^{J} \sum_{i=1}^{N} E_q  \left[ \log p \left( r_{ji} | \theta_{ji}, n_{ji} \right) \right] \\
&= \sum_{j=1}^{J} \sum_{i=1}^{N} \log \left( \frac{ \Gamma(n_{ji}+1) } { \Gamma(r_{ji}+1) \Gamma( n_{ji} - r_{ji} + 1 ) }\right)  \\
&\quad + \sum_{j=1}^{J} \sum_{i=1}^{N} \left\lbrace r_{ji} E_q \left[ \log \theta_{ji} \right] + (n_{ji} - r_{ji}) E_q  \left[  \log (1 - \theta_{ji}) \right] \right\rbrace
\end{split}
\end{equation}
\begin{equation}
\begin{split}
\label{mu}
E_q \left[ \log p\left(\mu ; \mu_0, M_0 \right)\right] &= \sum_{j=1}^{J} E_q  \left[ \log p\left( \mu_j; \mu_0, M_0 \right) \right] \\
&= J* \log \frac{ \Gamma(M_0) } { \Gamma(\mu_0 M_0) \Gamma(M_0 (1-\mu_0))} \\
&\quad + \sum_{j=1}^{J} \left\lbrace (M_0\mu_0 -1)E_q  \left[ \log \mu_j \right] + (M_0 ( 1 - \mu_0) - 1) E_q  \left[ \log (1 - \mu_j)\right]\right\rbrace
\end{split}
\end{equation}
\begin{equation}
\begin{split}
\label{theta}
E_q \left[ \log p\left(\theta | \mu; M \right)\right] &= \sum_{j=1}^{J} \sum_{i=1}^{N} E_q \left[ \log p\left(\theta_{ji} | \mu_j; M_j \right)\right] \\
&= N* \sum_{j=1}^{J} E_q  \left[ \log \left( \frac{ \Gamma(M_j) } { \Gamma(\mu_j M_j) \Gamma(M_j (1-\mu_j)) }\right) \right] \\
&\quad + \sum_{j=1}^{J} \sum_{i=1}^{N} \left\lbrace M_j E_q \left[ \mu_j \right] E_q \left[ \log \theta_{ji} \right] - E_q  \left[ \log \theta_{ji} \right] \right\rbrace\\
&\quad + \sum_{j=1}^{J} \sum_{i=1}^{N} \left\lbrace \left( M_j - 1 - M_j E_q\left[ \mu_j \right]  \right) E_q\left[ \log \left( 1 - \theta_{ji}\right) \right] \right\rbrace
\end{split}
\end{equation}

Therefore, we need to compute the following expectations with respect to the variational distribution:
$ E_q \left[ \log \theta_{ji} \right] $, $ E_q\left[ \log \left( 1 - \theta_{ji}\right) \right] $ , $ E_q  \left[ \log \mu_j \right] $ , $ E_q  \left[ \log (1 - \mu_j)\right] $, $ E_q \left[ \mu_j \right] $, and $ E_q\left[ \log \left( \frac{ \Gamma(M_j) } { \Gamma(\mu_j M_j) \Gamma(M_j (1-\mu_j)) }\right)\right] $.

We select the functional forms for the variational distributions $q(\theta)$ and $q(\mu)$ to facilitate these expected value computations.

\subsection{Variational Distributions}
Since $\theta$ and $r$ are conjugate pairs, the posterior distribution of $\theta_{ji}$ is a Beta distribution,
\begin{align}
&p(\theta_{ji}|r_{ji},n_{ji},\mu_j,M_j)
\thicksim \text{Beta}(r_{ji}+M_j \mu_j, n_{ji}-r_{ji}+M_j(1-\mu_j)).
\end{align}
Therefore, we propose a Beta distribution with parameter vector $\delta_{ji}$ as variational distribution,
\begin{align}
\theta_{ji} &\thicksim \text{Beta}(\delta_{ji1}, \delta_{ji2}) \nonumber.
\end{align}
The posterior distribution of $\mu_j$ is given by its Markov blanket,
\begin{align}
p(\mu_j|\theta_{ji},M_j,\mu_0,M_0)\propto p(\mu_j|\mu_0,M_0)p(\theta_{ji}|\mu_j,M_j).
\end{align}
This is not in the form of any known distribution.
But, since the support of $\mu_j$ is $[0,1]$, we propose a Beta distribution with parameter vector $\gamma_{j}$ as variational distribution,
\begin{align}
\mu_j &\thicksim \text{Beta}(\gamma_{j1}, \gamma_{j2}) \nonumber.
\end{align}
Given these variational distributions, we have
\begin{align}
E_q \left[ \log \theta_{ji} \right] &= \psi(\delta_{ji1}) - \psi(\delta_{ji1}+\delta_{ji2}) \nonumber \\
E_q \left[ \log \left( 1 - \theta_{ji}\right) \right]&= \psi(\delta_{ji2}) - \psi(\delta_{ji1}+\delta_{ji2}) \nonumber \\
E_q \left[ \mu_j \right] &= \frac{\gamma_{j1}}{\gamma_{j1} + \gamma_{j2}} \nonumber \\
E_q  \left[ \log \mu_j \right] &= \psi(\gamma_{j1}) - \psi(\gamma_{j1}+\gamma_{j2}) \nonumber \\
E_q  \left[ \log (1 - \mu_j)\right] &= \psi(\gamma_{j2}) - \psi(\gamma_{j1}+\gamma_{j2})\nonumber, \\
\end{align}
where $\psi$ is the digamma function.

Since there is no analytical representation for $ E_q\left[ \log \left( \frac{ \Gamma(M_j) } { \Gamma(\mu_j M_j) \Gamma(M_j (1-\mu_j)) }\right)\right] $, we must resort to numerical integration,

\begin{equation}\label{eqn:integration}
\begin{split}
E_q\left[ \log \left( \frac{ \Gamma(M_j) } { \Gamma(\mu_j M_j) \Gamma((1-\mu_j)M_j ) }\right)\right] &= \int_{0}^{1} q(\mu_j;\gamma_{j1}, \gamma_{j2}) \log \left( \frac{ \Gamma(M_j) } { \Gamma(\mu_j M_j) \Gamma((1-\mu_j)M_j ) }\right) d\mu_j.
\end{split}
\end{equation}
Here $q(\mu_j;\gamma_{j1}, \gamma_{j2})$ is the probability density function of the Beta distribution that is calculated using the Python built-in function \texttt{scipy.stats.beta.pdf},
and $\log \left( \frac{ \Gamma(M_j) } { \Gamma(\mu_j M_j) \Gamma((1-\mu_j)M_j ) }\right)$ is calculated using the Python built-in function \texttt{scipy.special.betaln}.
Unfortunately, this numerical integration step is computationally expensive.
Finally, the entropy terms can be computed as follows,
\begin{equation}
\begin{split}
E_q \left[ \log q\left(\mu \right)\right] &= \sum_{j=1}^{J} E_q \left[ \log q(\mu_j)\right] \\
&= -\sum_{j=1}^{J} \left\lbrace \log (B(\gamma_{j1},\gamma_{j2})) -(\gamma_{j1}-1)\psi(\gamma_{j1}) \right\rbrace \\
&\quad  + \sum_{j=1}^{J} \left\lbrace -(\gamma_{j2}-1)\psi(\gamma_{j2}) + (\gamma_{j1}+\gamma_{j2}-2)\psi(\gamma_{j1}+\gamma_{j2})\right\rbrace;
\end{split}
\end{equation}
and
\begin{equation}
\begin{split}
E_q \left[ \log q\left(\theta \right)\right] &= \sum_{j=1}^{J}\sum_{i=1}^{N} E_q\left[ \log q(\theta_{ji})\right] \\
&= -\sum_{j=1}^{J}\sum_{i=1}^{N} \left\lbrace \log (B(\delta_{ji1},\delta_{ji2}))-(\delta_{ji1}-1)\psi(\delta_{ji1}) \right\rbrace \\
&\quad + \sum_{j=1}^{J}\sum_{i=1}^{N} \left\lbrace -(\delta_{ji2}-1)\psi(\delta_{ji2}) + (\delta_{ji1}+\delta_{ji2}-2)\psi(\delta_{ji1}+\delta_{ji2})\right\rbrace.
\end{split}
\end{equation}
\subsection{Variational EM Algorithm}
Variational EM algorithm maximizes the ELBO on the true likelihood by alternating between maximization over $q$ (E-step) and maximization over $\phi= \left\{\mu_0, M_0, M \right\}$ (M-step).
\subsubsection{(E-step): Updating the variational distributions}
The terms in the ELBO that depend on $q(\theta_{ji} | \delta_{ji1}, \delta_{ji2})$ are
\begin{equation}\label{eqn:partial_theta}
\begin{split}
\mathcal{L}_{{[q(\theta_{ji})]}}
& = \sum_{j=1}^{J} \sum_{i=1}^{N} \left\lbrace r_{ji} E_q \left[ \log \theta_{ji} \right] + (n_{ji} - r_{ji}) E_q  \left[  \log (1 - \theta_{ji}) \right] \right\rbrace\\
&\quad  +  \sum_{j=1}^{J} \sum_{i=1}^{N} \left\lbrace M_j E_q \left[ \mu_j \right] E_q \left[ \log \theta_{ji} \right] - E_q  \left[ \log \theta_{ji} \right] + \left( M_j - 1 - M_j E_q\left[ \mu_j \right]  \right) E_q\left[ \log \left( 1 - \theta_{ji}\right) \right] \right\rbrace\\
&\quad - \sum_{j=1}^{J}\sum_{i=1}^{N} E_q\left[ \log q(\theta_{ji})\right]
\end{split}
\end{equation}

We update the variational parameters by numerically optimizing $\hat{\delta}_{ji1}, \hat{\delta}_{ji2} = \arg \max_{\delta_{ji1}, \delta_{ji2}} \mathcal{L}_{{[q(\theta_{ji})]}}$ subject to the constraints that $\delta_{ji1} \geq 0$ and $\delta_{ji2} \geq 0$ and conditioned on fixed values for the other model and variational parameters using Sequential Least SQuares Programming (SLSQP) ~\citep{kraft1988software}.

We update the variational distribution $q(\mu_j)$ using the partial ELBO depending on $\gamma_j$ from each position $j$ \eqref{eqn:partial_mu}.
\begin{equation}\label{eqn:partial_mu}
\begin{split}
\mathcal{L}_{{[q(\mu_j)]}}
& = N \sum_{j=1}^{J} E_q  \left[ \log \left( \frac{ \Gamma(M_j) } { \Gamma(\mu_j M_j) \Gamma(M_j (1-\mu_j)) }\right) \right] \\
&\quad + \sum_{j=1}^{J} \sum_{i=1}^{N} \left\lbrace M_j E_q \left[ \mu_j \right] E_q \left[ \log \theta_{ji} \right] - E_q  \left[ \log \theta_{ji} \right] + \left( M_j - 1 - M_j E_q\left[ \mu_j \right]  \right) E_q\left[ \log \left( 1 - \theta_{ji}\right) \right] \right\rbrace\\
&\quad + J \log \frac{ \Gamma(M_0) } { \Gamma(\mu_0 M_0) \Gamma(M_0 (1-\mu_0))} \\
&\quad + \sum_{j=1}^{J} \left\lbrace (M_0\mu_0 -1)E_q  \left[ \log \mu_j \right] + (M_0 ( 1 - \mu_0) - 1) E_q  \left[ \log (1 - \mu_j)\right]\right\rbrace\\
&\quad - \sum_{j=1}^{J} E_q \left[ \log q(\mu_j)\right]
\end{split}
\end{equation}
Again, we update the variational parameters by numerically optimizing $\hat{\gamma}_{j1}, \hat{\gamma}_{j2} = \arg \max_{\gamma_{j1}, \gamma_{j2}} \mathcal{L}_{{[q(\mu_{j})]}}$ subject to the constraints that $\gamma_{j1} \geq 0$ and $\gamma_{j2} \geq 0$ and conditioned on fixed values for the other model and variational parameters using SLSQP ~\citep{kraft1988software}.
The computational cost of optimizing \eqref{eqn:partial_mu} is high because of the quadrature of $E_q\left[ \log \left( \frac{ \Gamma(M_j) } { \Gamma(\mu_j M_j) \Gamma(M_j (1-\mu_j)) }\right)\right]$ in \eqref{eqn:integration}.

\subsubsection{(M-step): Updating the model parameters}
We can write out the ELBO as a function of each model parameter $\mu_0$, $M_0$, and $M_j$ as follows.

The ELBO with respect to $ \mu_0 $ is
\begin{equation}\label{eqn:mu_0}
\begin{split}
\mathcal{L}_{[\mu_0]}
&= -J*\log  \Gamma(\mu_0 M_0) - J*\log \Gamma(M_0 (1-\mu_0))
+ M_0\mu_0\sum_{j=1}^{J} \left\lbrace E_q  \left[ \log \mu_j \right]
- E_q  \left[ \log (1 - \mu_j)\right]\right\rbrace . \\
\end{split}
\end{equation}

The ELBO with respect to $ M_0 $ is
\begin{equation}\label{eqn:M_0}
\begin{split}
\mathcal{L}_{[M_0]}
&=J* \log \frac{ \Gamma(M_0) } { \Gamma(\mu_0 M_0) \Gamma(M_0 (1-\mu_0))}
+ M_0 \sum_{j=1}^{J} \left\lbrace \mu_0E_q  \left[ \log \mu_j \right] + ( 1 - \mu_0) E_q  \left[ \log (1 - \mu_j)\right]\right\rbrace.  \\
\end{split}
\end{equation}

The ELBO with respect to $M_j$ is
\begin{equation}\label{eqn:M}
\begin{split}
\mathcal{L}_{{[M_j]}}
&= N* \sum_{j=1}^{J} E_q  \left[ \log \left( \frac{ \Gamma(M_j) } { \Gamma(\mu_j M_j) \Gamma(M_j (1-\mu_j)) }\right) \right] \\
&\quad + M_j \sum_{j=1}^{J} \sum_{i=1}^{N} \left\lbrace E_q \left[ \mu_j \right] E_q \left[ \log \theta_{ji} \right] + \left( 1 - E_q\left[ \mu_j \right]  \right) E_q\left[ \log \left( 1 - \theta_{ji}\right) \right] \right\rbrace. \\
\end{split}
\end{equation}
We also use SLSQP to optimize the ELBO function with respect to each parameter, $\mu_0$, $M_0$, and $M_j$.
It is computationally easy to optimize $\mu_0$ \eqref{eqn:mu_0} and $M_0$ \eqref{eqn:M_0}.
However, it is costly for optimizing $M_j$ \eqref{eqn:M} because the quadrature is needed to calculate $ E_q\left[ \log \left( \frac{ \Gamma(M_j) } { \Gamma(\mu_j M_j) \Gamma(M_j (1-\mu_j)) }\right)\right] $ using \eqref{eqn:integration}.

The variational EM algorithm is summarized using pseudocode in Algorithm 1.
\begin{algorithm}[ht]
  \caption{Variational EM Inference}

  \begin{algorithmic}[1]

  \State Initialize $q(\theta, \mu)$ and $\hat{\phi}$

  \State // M-step

  \Repeat

    \State // E-step

	\Repeat

		\For {j = 1 to J}
			\For {i = 1 to N}
			\State Optimize $\mathcal{L}(q, \hat{\phi})$ over $q(\theta_{ji}; \delta_{ji}) = \text{Beta} (\delta_{ji})$
			\EndFor
		\EndFor

        \For {j = 1 to J}
            \State Optimize $\mathcal{L}(q, \hat{\phi})$ over $q(\mu_j; \gamma_j) = \text{Beta} (\gamma_j)$
        \EndFor

    \Until{change in $\mathcal{L}(q,\hat{\phi})$ is small}

    \State Set $\hat{\phi} \leftarrow \arg \max\limits_{\phi}
            \mathcal{L}(\hat{q},\phi)$
  \Until {change in $\mathcal{L}(\hat{q},\phi)$ is small}

  \end{algorithmic}

\end{algorithm}
\section{Hypothesis Testing}
The posterior distribution over $\mu_j^{\triangle} \mid r^{case}, r^{control} \triangleq \mu_j|r^{case} - \mu_j|r^{control}$ is the distribution over the change in the non-reference read rate at position $j$ between a case and control sample.
However, we cannot compute the posterior distribution, $\mu_j^{\triangle} | r^{case}, r^{control}$, because we do not have access to the component posterior distributions $\mu_j|r^{case}$ and $\mu_j|r^{case}$.
But, by approximating the distribution of $\mu_j|r^{case}$ and $\mu_j|r^{control}$ as Gaussians with moments matched to the variational posterior distributions, the distribution of $\mu_j^{\triangle} | r^{case}, r^{control}$ can be approximated as a Gaussian.

Under the variational approximation,
\begin{align}
  E_q[\mu_j|r^{case}] &= \frac{\gamma_{j1}^{case}}{\gamma_{j1}^{case} + \gamma_{j2}^{case}}
  \\
  \text{Var}_q[\mu_j|r^{case}] &= \frac{\gamma_{j1}^{case} \gamma_{j2}^{case}}{(\gamma_{j1}^{case} + \gamma_{j2}^{case} + 1)(\gamma_{j1}^{case} + \gamma_{j2}^{case})^2}
\end{align}
for $\mu_j|r^{case}$ and likewise for $\mu_j|r^{control}$.
We approximate the posterior for the case sample as
\begin{equation}
  \mu_j | r^{case} \sim \mathcal{N}(E_q[\mu_j|r^{case}], \text{Var}_q[\mu_j|r^{case}])
\end{equation}
and likewise for the control.
Then,
\begin{equation}
  \mu_j^{\triangle} \mid r^{case}, r^{control} \sim \mathcal{N}(E_q[\mu_j|r^{case}] - E_q[\mu_j|r^{control}], \text{Var}_q[\mu_j|r^{case}] + \text{Var}_q[\mu_j|r^{control}])
\end{equation}

Now, we can approximate the posterior probability of interest,
\begin{equation}
  \Pr( \mu_j^{\triangle} \geq \tau \mid r^{case}, r^{control} ),
\end{equation}
that is, the posterior probability that the difference in the non-reference read rate is greater than a fixed effect size $\tau$ (e.g. zero) for a one sided test.
For a two sided test, we compute the approximate probability
\begin{equation}
  \Pr( | \mu_j^{\triangle} | \geq \tau \mid r^{case}, r^{control}).
\end{equation}
A position is called a \textit{provisional variant} if $\Pr( | \mu_j^{\triangle} | \geq \tau \mid r^{case}, r^{control}) \geq 1-\alpha/2$, where the probability is approximated as described.

It is possible that a position is called a variant due to a differential non-reference read count, but no particular alternative base is more frequently observed than the others.
In this case, the likely case is a sequencing error that indiscriminately incorporates a non-reference base at the position.
To discriminate this non-biological cause from the interesting true variants we use a $\chi^2$ goodness-of-fit test for non-uniform base distribution~\citep{efron2010large, he2015rvd2}.
For each provisional variant, if we reject the null hypothesis that the distribution is uniform, we promote the position to a called variant.

\section{Results}

\subsection{Data Sets}

\subsubsection{Synthetic DNA Sequence Data}

Two random 400bp DNA sequences were synthesized. One sample with 14 loci variants is taken as the case and the other without variants is taken as the control.
Case and control samples were mixed to yield defined NRAF of 0.1\%, 0.3\%, 1.0\%, 10.0\%, and 100.0\%.
The details of the experimental protocol are available in \citet{flaherty2011ultrasensitive}.
The synthetic DNA data were downsampled by $10\times$, $100\times$, $1,000\times$, and $10,000\times$ using \texttt{picard} (v 1.96).
The final data set contains read pairs for six replicates for the control and cases at different NRAF levels.

\subsubsection{Longitudinal Directed Evolution Data}

The longitudinal yeast data comes from three strains of haploid S288c which were grown for 448 generations under limited-glucose (0.08\%).
The wild-type ancestral strain GSY1136 was sequenced as a reference.
Aliquots were taken about every 70 generations and sequenced.
The detail of library sequencing is described in \citet{kvitek2013whole}, \citet{bansal2010statistical}, and \citet{kao2008molecular}.
The Illumina sequencing data are available on the NCBI Sequence Read Archive (SRA054922)~\citep{kvitek2013whole}.
For this study, we received the original BAM files from one of the authors.
The aligned BAM files have $266$ - $1,046\times$ coverage.
We used \texttt{samtools} (v 1.1) with \texttt{-mpileup -C50} flags to convert BAM files to pileup files.
Then, we generated depth chart files, which are tab-delimited text files recording the count of the number of nucleotides in columns and genomic positions in rows.
We ran our variational inference algorithm on the depth chart files to identify SNVs.

\subsection{Performance on Synthetic DNA Data}

\subsubsection{Comparison of Sensitivity and Specificity}

We compared the performance of our variational EM algorithm and the MCMC sampling algorithm in terms of sensitivity and specificity (Table~\ref{tbl:statistics_mcmc_var}).
We used the posterior distribution test and $\chi^2$ test to detect variants for a broad range of median read depths and NRAFs.
The variational EM algorithm shows higher sensitivity and specificity than the MCMC algorithm in the events when NRAF is 0.1\%.
The variational EM algorithm has a higher specificity compared with the MCMC algorithm for a median read depth of $41,472\times$ at 0.3\% NRAF and $55,489\times$ at 1.0\% NRAF, but the sensitivity is slightly lower due to false negatives.
%
\begin{table}[ht]
\centering
\includegraphics[width=0.7\textwidth]{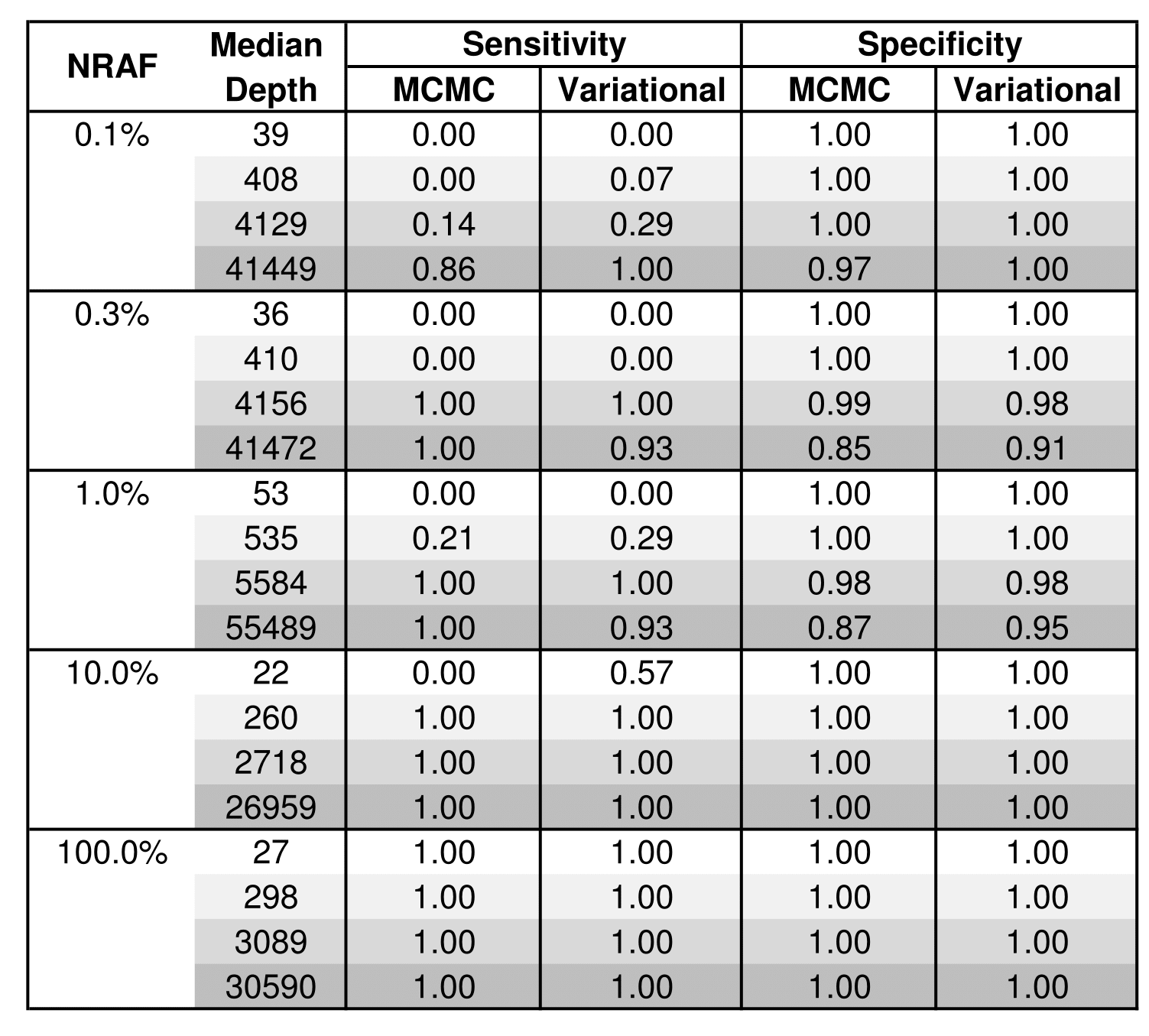}
\vspace{5pt}
\caption{Sensitivity/Specificity comparison of variational EM algorithm with the MCMC algorithm on the synthetic DNA data set.}
\label{tbl:statistics_mcmc_var}
\end{table}

\subsubsection{Comparison of Approximated Posterior Distribution}

\begin{figure}[htbp]
\centering
\includegraphics[width=0.7\textwidth]{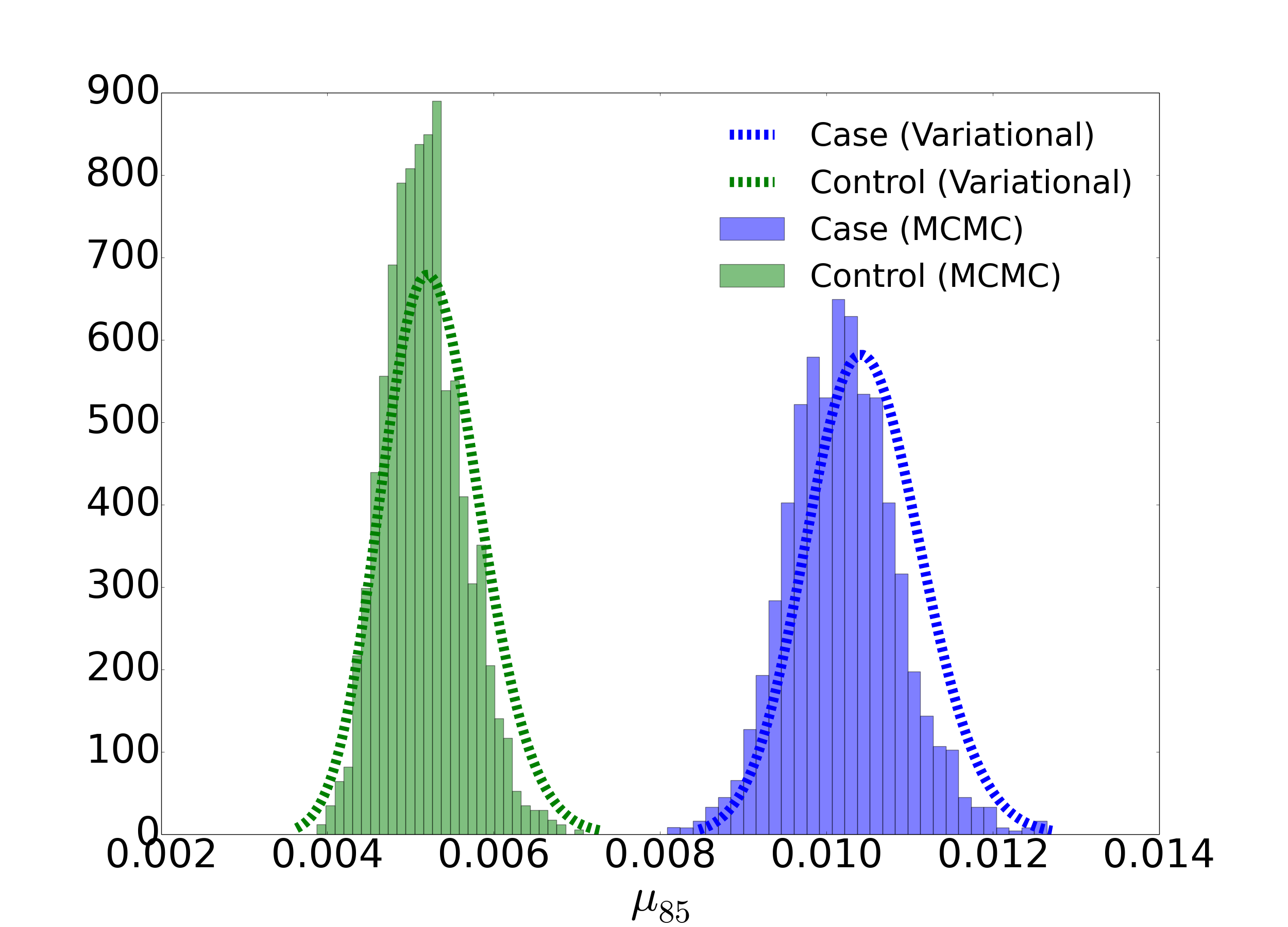}
\caption{Approximated posterior distributions by the variational EM and MCMC algorithms on a true variant position 85 when the median read depth is $5,584\times$.}
\label{tbl:compare1}
\end{figure}

Figure~\ref{tbl:compare1} shows the approximate posterior distribution of the variational EM algorithm and samples of the MCMC algorithm.
One variant position, 85, is taken as an example to show the comparison of the approximated posteriors.
The variational EM and MCMC algorithms both identify all the variants when NRAF is 10.0\% and 100.0\%.
The variational EM algorithm calls 90 false positive positions without a $\chi^2$ test when NRAFs are 0.1\% and 0.3\% for low median read depth ($30\times$ and $400\times$).
This is to be expected because it is highly unlikely to correctly identify a variant base with a population frequency of $1$ in $1,000$ with less than a $1,000\times$ read depth.

\begin{figure}[htbp]
\centering
\includegraphics[width=0.7\textwidth]{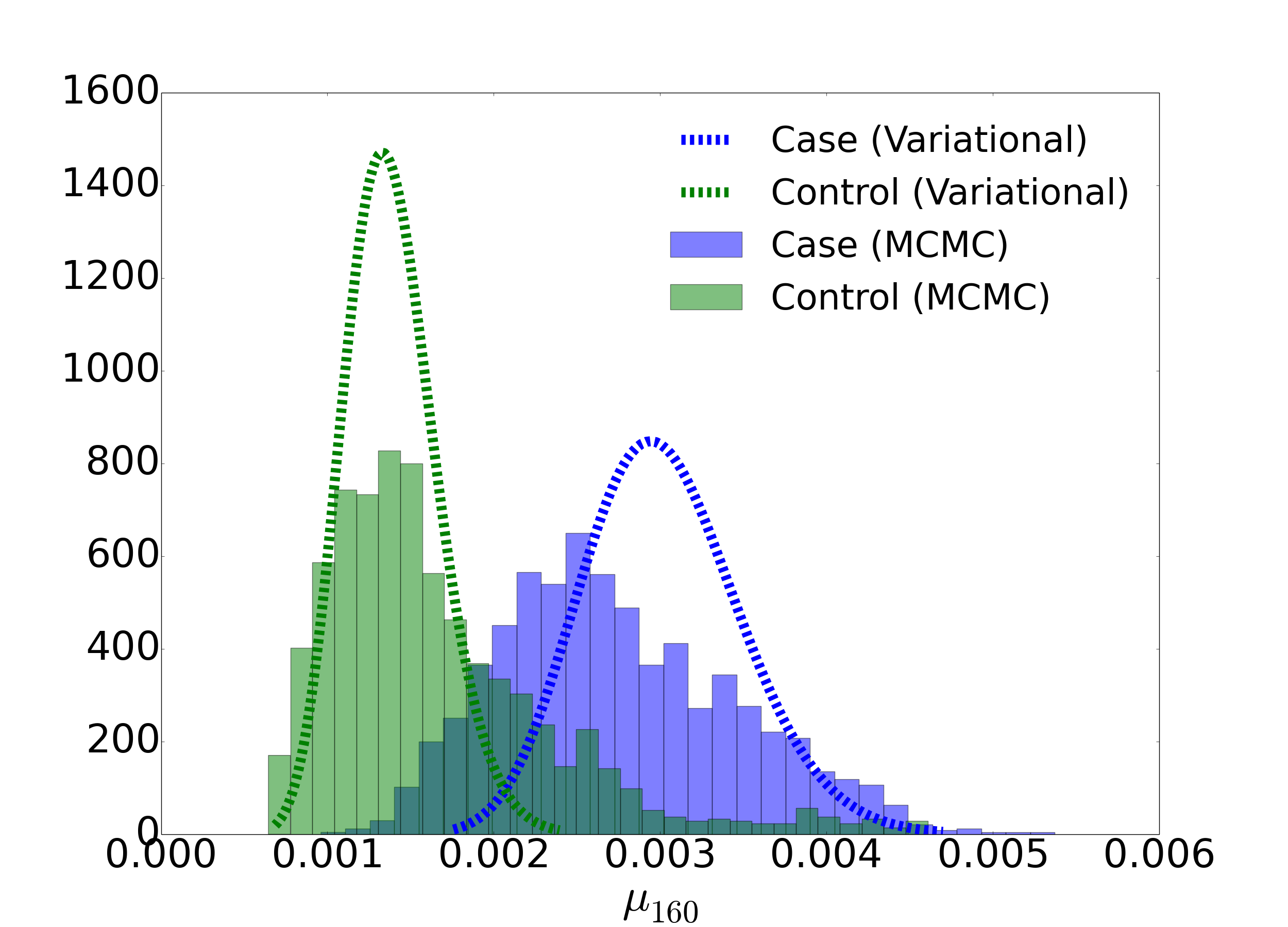}
\caption{Approximated posterior distribution by the variational EM and MCMC algorithms on a false variant position 160 when the median read depth is $410\times$.}
\label{tbl:compare2}
\end{figure}

A false positive, a non-mutated position that is called by the variational EM algorithm but not called by the MCMC algorithm, is shown in Figure~\ref{tbl:compare2}.
The variance of MCMC posterior estimate is higher than that of the variational posterior estimate.
We tested 10 random initial values variational inference algorithm and found the approximate posterior distributions from the variational EM algorithm are essentially equivalent for all random initializations.
It is notable that the shape of the proposed Beta variational distribution is well approximated by a Gaussian.

\subsubsection{Comparison to the State-of-the-Art Methods}
We compared the performance of our variational EM algorithm with the state-of-the-art variant detection methods, SAMtools ~\citep{li2009sequence}, GATK ~\citep{mckenna2010genome}, VarScan2 ~\citep{koboldt2012varscan}, Strelka ~\citep{saunders2012strelka}, MuTect ~\citep{cibulskis2013sensitive}, and RVD2 ~\citep{he2015rvd2}, using synthetic DNA data set (Table~\ref{tbl:character_all}).
Among all of the methods compared, our variational EM algorithm has a higher sensitivity and specificity for a broad range of read depths and NRAFs.
Our variational EM algorithm shows higher specificity than all the other tested methods at a very low NRAF (0.1\%) level.
However, our algorithm has a slightly lower specificity than the MCMC algorithm when the median read depth is $4,156\times$ at 0.3\% NRAF, and a slightly lower sensitivity than the MCMC algorithm when the median read depth is $41,472\times$ at 0.3\% NRAF and a median read depth of $55,489\times$ at 1.0\% NRAF.
The performance of other methods is stated in detail in \citet{he2015rvd2}.
\begin{table*}[htbp]
\centering
\includegraphics[width=1.0\textwidth]{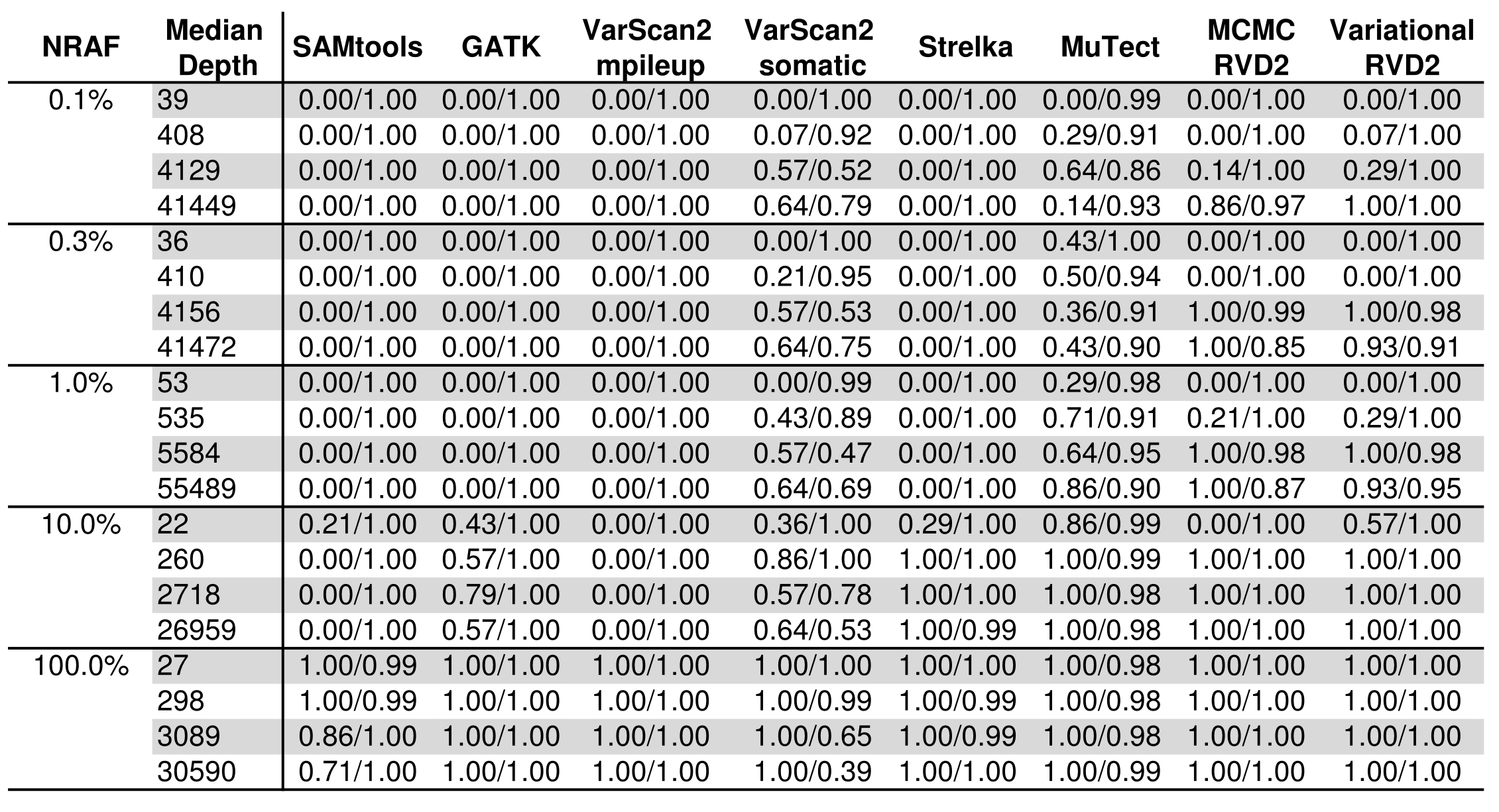}
\caption{Sensitivity/Specificity comparison of our variational RVD2 with other variant detection methods on synthetic DNA data.}
\label{tbl:character_all}
\end{table*}

\subsubsection{Comparison of Timing}
\begin{figure}[ht]
\centering
\includegraphics[width=0.6\textwidth]{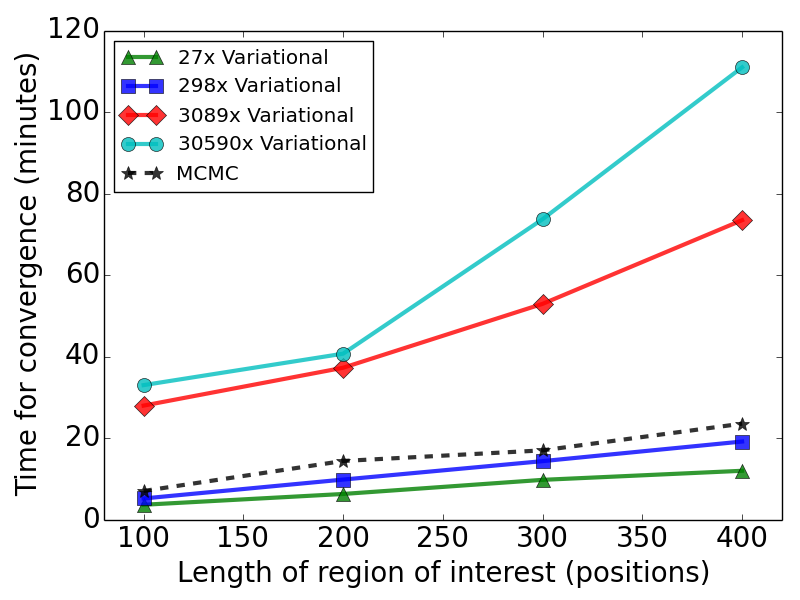}
\caption{Timing for our variational EM algorithm and MCMC sampling algorithm.
Sixty processers are used to estimate the model on the synthetic data set.}
\label{tbl:timing_mcmc_var}
\end{figure}

The computational time for approximating the variational posterior distribution is increased by expanding the length of region and the median read depth (Figure~\ref{tbl:timing_mcmc_var}).
Our variational EM algorithm is faster than the MCMC algorithm at the low median read depths of $27\times$ and $298\times$, and slower for the high median read depths of $3,089\times$ and $30,590\times$.

Table~\ref{tbl:timing_profile_all} shows the timing profile for each part of our variational EM algorithm when median read depth is $3,089\times$.
Optimizing $\gamma$ in the E-step \eqref{eqn:partial_mu} and optimizing $M_j$ in the M-step \eqref{eqn:M} takes more than 95\% of the time of one variational iteration in a test of a single processor, since the integration \eqref{eqn:integration} is needed.
\begin{table*}[htbp]
\centering
\includegraphics[width=1.0\textwidth]{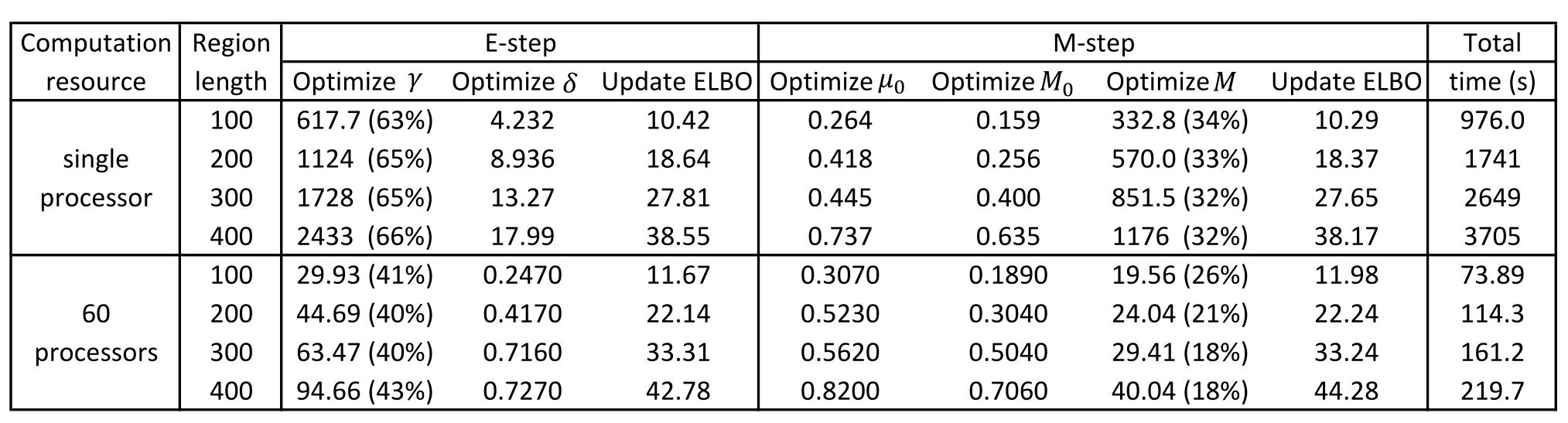}
\caption{Timing profile of 4 significant figures for one iteration of variational EM algorithm when median read depth is $3,089\times$.
Single and multiple processors are both tested to estimate timing. Time for optimizing $\gamma$ in the E-step and optimizing $M$ in the M-step is highlighted in percentage.}
\label{tbl:timing_profile_all}
\end{table*}
\subsection{Variant Detection on the Longitudinal Directed Evolution Data}
We applied our variational EM algorithm to the MTH1 gene at Chr04:1,014,401-1,015,702 (1,302bp), which is the most frequently observed mutated gene by \citet{kvitek2013whole}.
Our algorithm detected the same variants that were found by \citet{kvitek2013whole} (shown as highlighted in Table~\ref{tbl:mutations}).
Additionally, we detected 81 novel variants in 8 timepoints that the original publication did not detect.
In Table~\ref{tbl:mutations}, G7 is the baseline NRAF as the control sample when comparing with G70, G133, G266, G322, G385, and G448 in the respective hypotheses testing.
The corresponding NRAFs of called variants at different time points are given by the estimate of the latent variable, $\hat{\mu_j} = E_q[\mu_j|r]$.
\begin{table*}[htbp]
\centering
\includegraphics[width=1.0\textwidth]{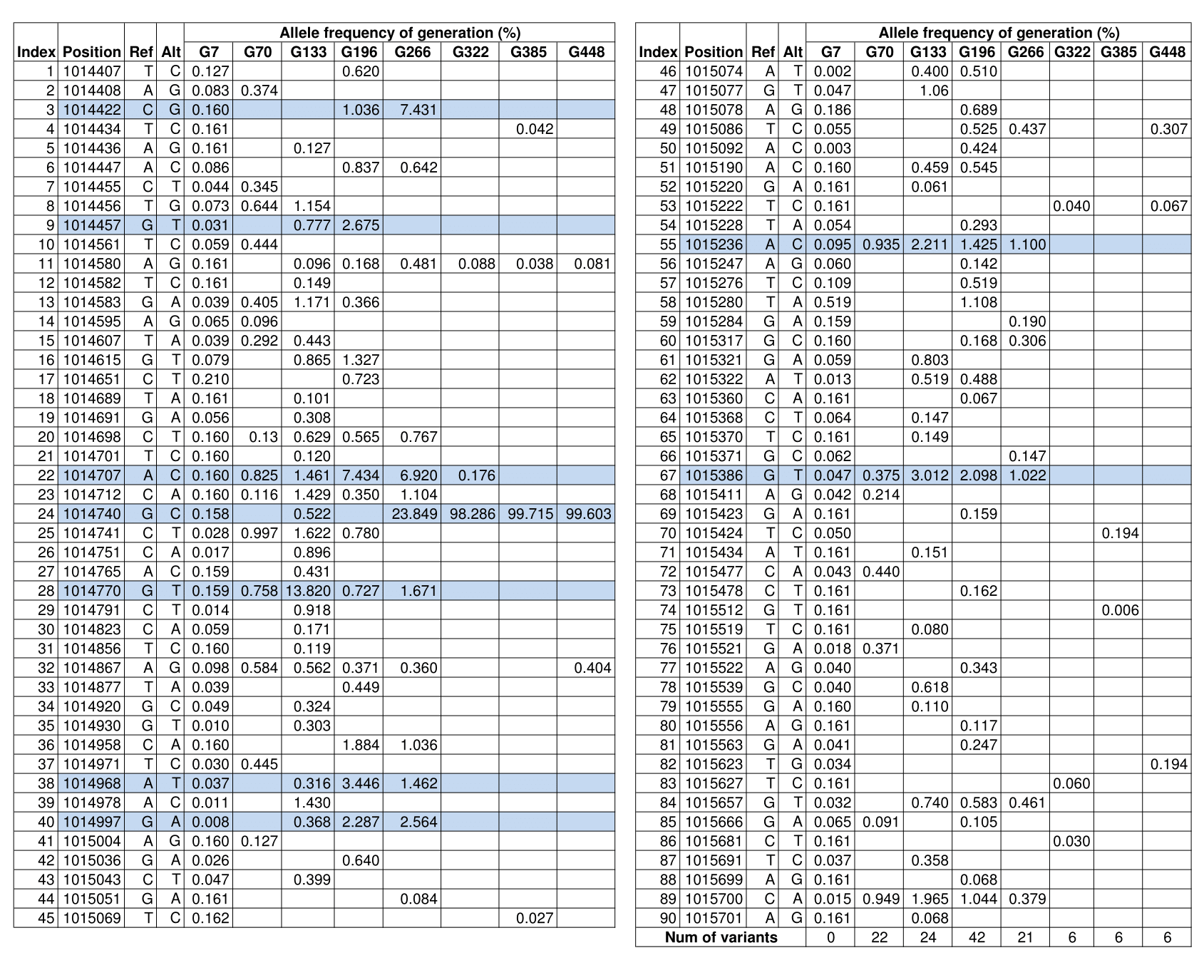}
\caption{Identified variants and corresponding NRAFs in gene MTH1 on Chromosome 4.
A blank cell indicates that the position of that time point is not called significantly different than G7.
The positions highlighted as blue were also identified by Kvitek, 2013.
The other 81 positions are novel identified variants in 8 timepoints.}
\label{tbl:mutations}
\end{table*}

All of these variants, except the variant at position Chr04:1,014,740, decrease in NRAF following a maximum.
The allele at position Chr04:1,014,740 is a beneficial variant that arises in NRAF to 99.603\% at generation 448 within a constant glucose-limited environment.
Moreover, we identified the first emergence of this beneficial variant as early as 0.522\% in generation 133.
Table~\ref{tbl:mutations} shows that we are able to detect many variants (NRAF $<$ 1.0\%) early in the evolutionary time course.
Furthermore, we detected a variant at position Chr04:1,015,666 at generation 70 as 0.091\% NRAF, which is less than a 0.1\% fraction.

\begin{figure}[htbp]
\centering
\includegraphics[width=0.6\textwidth]{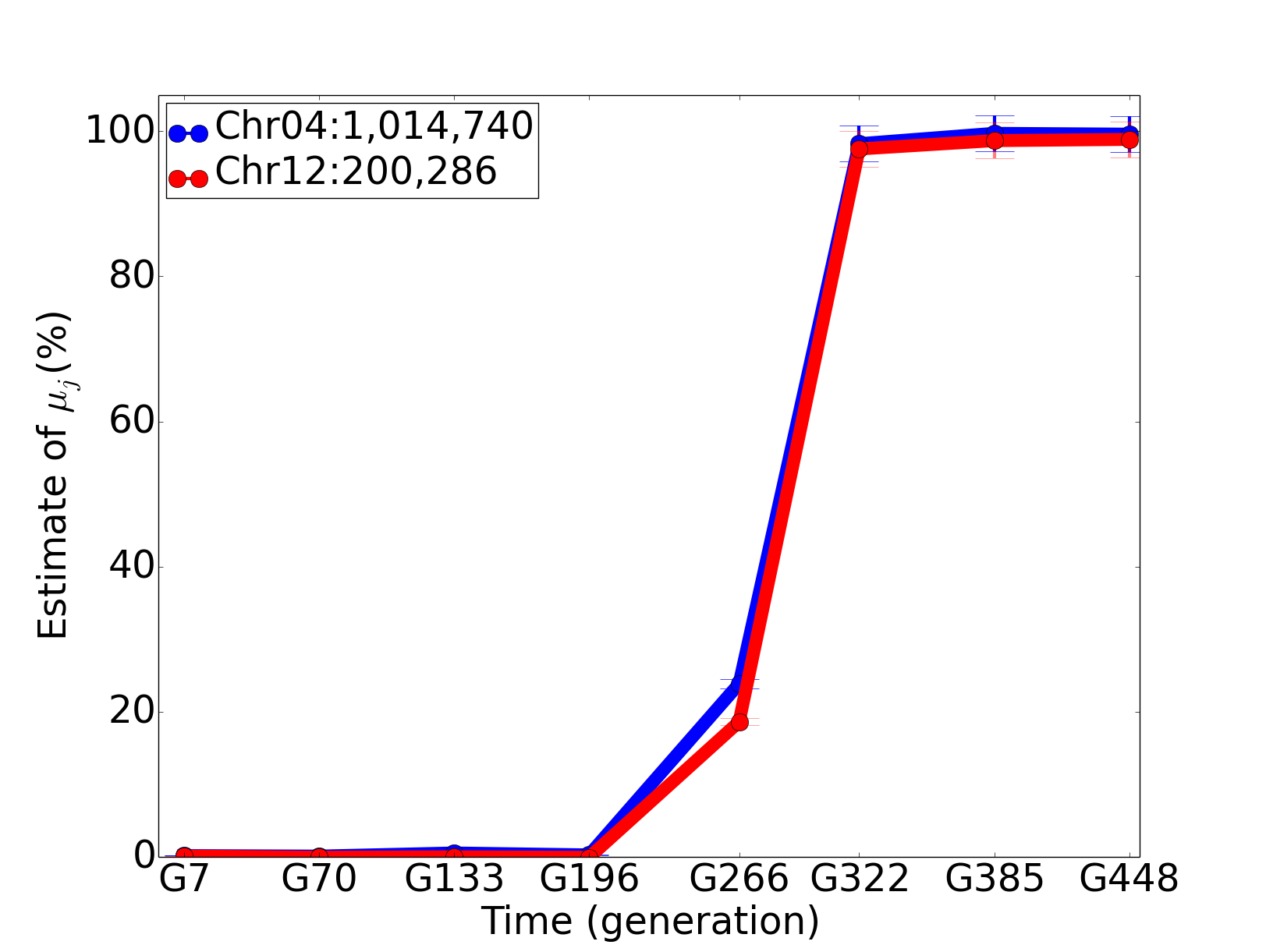}
\caption{The NRAF trend of concomitant variants in gene MTH1 and ADE16.
The 95\% Bayesian credible intervals are shown.}
\label{tbl:concomitant}
\end{figure}

We identified a pair of variants, Chr04:1,014,740 in gene MTH1 and Chr12:200,286 in gene ADE16, that increase in NRAF together in time (Figure~\ref{tbl:concomitant}).
We hypotheses that the variants are concomitant in the same strain.
In this pair of genes, gene MTH1 is a negative regulator of the glucose-sensing signal transduction pathway, and gene ADE16 is an enzyme of $\mathit{de\, novo}$ purine biosynthesis.
Glucose sensing induces gene expression changes to help yeast receive necessary nutrients, which could be a reason for this pair of genes to mutate together ~\citep{johnston1999feasting}.
Further experimental validation of this hypothesis would be required to definitively show that the mutations are concomitant.

\begin{figure}[htbp]
\centering
\includegraphics[width=0.8\textwidth]{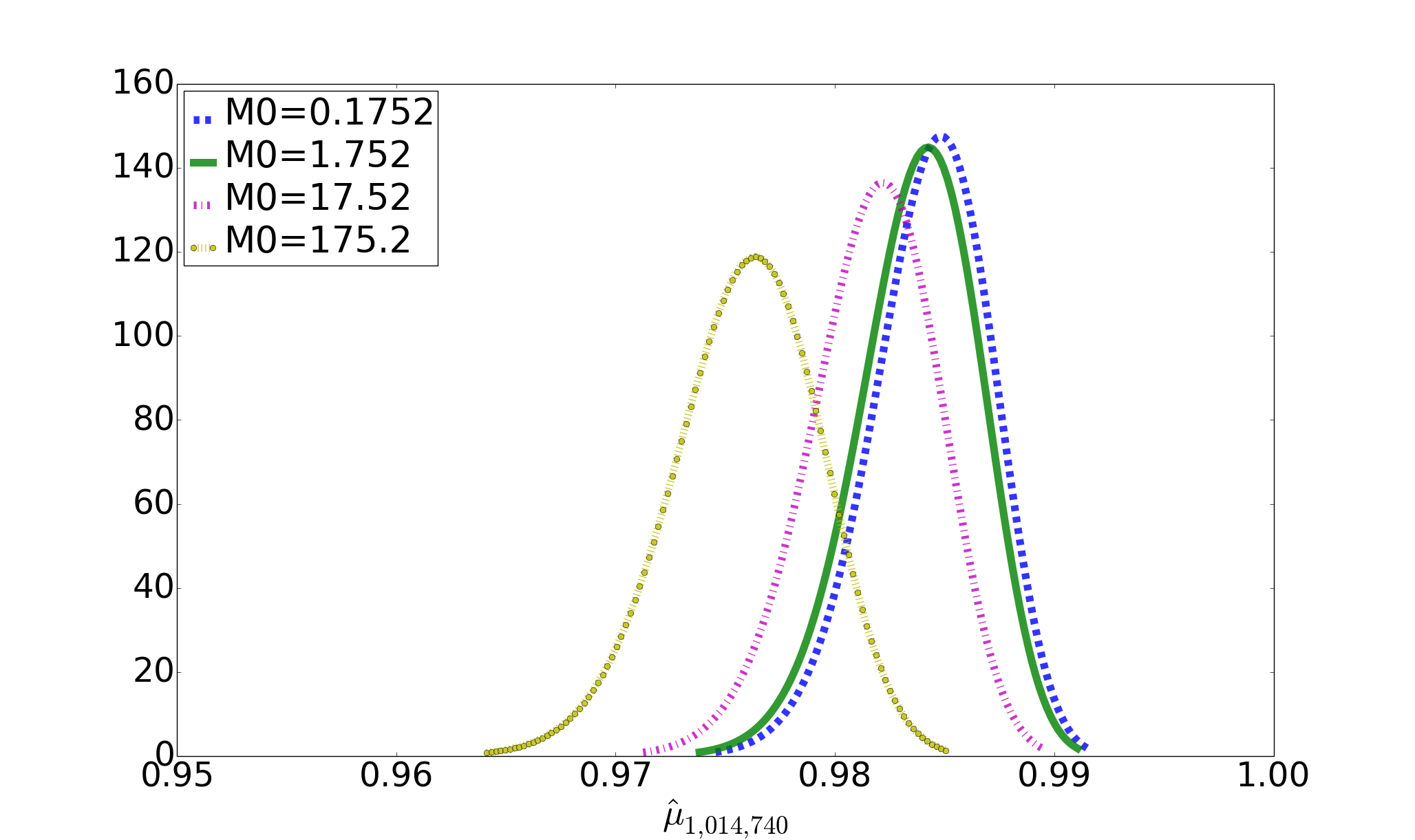}
\caption{Influence of $M_0$ on the estimate of $\mu_j$.
Posterior distributions of the variant at position Chr04:1,014,740, $\hat{\mu}_{1,014,740}$, with different $\hat{M_0}$ are shown.}
\label{tbl:M0}
\end{figure}

The global precision parameter $\hat{M_0}$ could influence the estimate of $\mu_j$ due to its regularization effect.
We show the influence of different $\hat{M_0}$ on variant position Chr04:1,014,740, $q(\mu_{1,014,740}|r)$ in Figure~\ref{tbl:M0}.
We see that as we decrease the prior precision parameter $\hat{M_0}$, $\hat{\mu}_{1,014,740}$ increases as expected.
But the effect of changing $\hat{M_0}$ over several orders of magnitude does not change $\mu_j$ greatly.
Here $\hat{M_0} = 1.752$ in this dataset.

\section{Discussion}
In this article, we propose a variational EM algorithm to estimate the non-reference allele frequency in the RVD2 model to identify rare nucleotide variants in heterogeneous pools.
We derived a variational EM algorithm to avoid an intractable integration in our Bayesian model.
Our results show that the variational EM algorithm
(i) is able to identify rare variants at a 0.1\% NRAF level with comparable sensitivity and specificity to a MCMC sampling algorithm;
(ii) has a higher specificity in comparison with many state-of-the-art algorithms in a broad range of NRAFs;
and (iii) detects SNVs early in the evolutionary time course, as well as tracks NRAF in a real longitudinal yeast data set.

We have chosen parametric forms for the variational distributions.
This choice has left us with a complex integral in our variational optimization problem.
In future work, we plan to explore other approximations of the variational distributions that render the integral easier to compute.
One could use cubic splines to numerically approximate the function and then integrate that surrogate ~\citep{mckinley1998cubic}.
Another strategy is to consider a Laplace approximation for the variational distribution, as we and others have done previously ~\citep{saddiki2014glad, wang2013variational}.


Improving the speed of the estimating algorithm enables us to interrogate whole-genome sequencing data.
By doing this, we hope to reveal the dynamics of arising variants at the genome-wide scale to show the genetic basis of clonal interference.
Our method can also be extended to study drug resistance by characterizing tumor heterogeneity in targeted anti-cancer chemotherapy samples.
We will be able to find the causative variants that lead to drug resistance and understand the causes of resistance at the single nucleotide level.

\section{Acknowledgments}
F.Z. and P.F. were supported by PhRMA Foundation Informatics Grant 2013080079.
We would like to thank Dan Kvitek for kindly sharing the data set used for the longitudinal yeast evolution analysis.
We acknowledge Chuangqi Wang and Ross Lagoy for valuable comments on the manuscript.
\appendix
\section{Derivation of the components in ELBO}
\label{appendix:ELBO}
Here we derive the requisite expectations with respect to the variational distribution for the ELBO.
Recall that $\theta_{ji} \thicksim \text{Beta}(\mu_j, M_j)$.
Then we can write,
\begin{equation}
\begin{split}
E_q \left[ \log p\left(\theta | \mu; M \right)\right] &= \sum_{j=1}^{J} \sum_{i=1}^{N} E_q \left[ \log p\left(\theta_{ji} | \mu_j; M_j \right)\right] \\
&= \sum_{j=1}^{J} \sum_{i=1}^{N}  E_q  \left[ \log \left( \frac{ \Gamma(M_j) } { \Gamma(\mu_j M_j) \Gamma(M_j (1-\mu_j)) } \theta_{ji}^{M_j\mu_j -1} (1 - \theta_{ji})^{M_j ( 1 - \mu_j) - 1} \right) \right] \\
&= \sum_{j=1}^{J} \sum_{i=1}^{N} E_q  \left[ \log \left( \frac{ \Gamma(M_j) } { \Gamma(\mu_j M_j) \Gamma(M_j (1-\mu_j)) }\right) \right] \\
&\quad + \sum_{j=1}^{J} \sum_{i=1}^{N}  E_q  \left[ \log \left( \theta_{ji}^{M_j\mu_j -1} (1 - \theta_{ji})^{M_j ( 1 - \mu_j) - 1} \right) \right] \\
&= \sum_{j=1}^{J} \sum_{i=1}^{N} E_q  \left[ \log \left( \frac{ \Gamma(M_j) } { \Gamma(\mu_j M_j) \Gamma(M_j (1-\mu_j)) }\right) \right]  \\
&\quad + \sum_{j=1}^{J} \sum_{i=1}^{N} \left\lbrace E_q \left[ \left( M_j\mu_j -1 \right) \log \theta_{ji} \right] + E_q \left[ \left( M_j ( 1 - \mu_j) - 1 \right) \log \left( 1 - \theta_{ji} \right) \right]\right\rbrace \\
&= \sum_{j=1}^{J} \sum_{i=1}^{N} E_q  \left[ \log \left( \frac{ \Gamma(M_j) } { \Gamma(\mu_j M_j) \Gamma(M_j (1-\mu_j)) }\right) \right] \\
&\quad + \sum_{j=1}^{J} \sum_{i=1}^{N} \left\lbrace M_j E_q \left[ \mu_j \right] E_q \left[ \log \theta_{ji} \right] - E_q  \left[ \log \theta_{ji} \right] \right\rbrace\\
&\quad + \sum_{j=1}^{J} \sum_{i=1}^{N} \left\lbrace \left( M_j - 1 - M_j E_q\left[ \mu_j \right]  \right) E_q\left[ \log \left( 1 - \theta_{ji}\right) \right] \right\rbrace\\
&= N* \sum_{j=1}^{J} E_q  \left[ \log \left( \frac{ \Gamma(M_j) } { \Gamma(\mu_j M_j) \Gamma(M_j (1-\mu_j)) }\right) \right] \\
&\quad + \sum_{j=1}^{J} \sum_{i=1}^{N} \left\lbrace M_j E_q \left[ \mu_j \right] E_q \left[ \log \theta_{ji} \right] - E_q  \left[ \log \theta_{ji} \right] \right\rbrace\\
&\quad + \sum_{j=1}^{J} \sum_{i=1}^{N} \left\lbrace \left( M_j - 1 - M_j E_q\left[ \mu_j \right]  \right) E_q\left[ \log \left( 1 - \theta_{ji}\right) \right] \right\rbrace
\end{split}
\end{equation}
Since $\mu_j \thicksim \text{Beta}(\mu_0, M_0)$,
\begin{equation}
\begin{split}
E_q \left[ \log p\left(\mu ; \mu_0, M_0 \right)\right] &= \sum_{j=1}^{J} E_q  \left[ \log p\left( \mu_j; \mu_0, M_0 \right) \right] \\
&= \sum_{j=1}^{J} E_q  \left[ \log \left( \frac{ \Gamma(M_0) } { \Gamma(\mu_0 M_0) \Gamma(M_0 (1-\mu_0)) } \mu_j^{M_0\mu_0 -1} (1 - \mu_j)^{M_0 ( 1 - \mu_0) - 1} \right) \right] \\
&= J* \log \frac{ \Gamma(M_0) } { \Gamma(\mu_0 M_0) \Gamma(M_0 (1-\mu_0))} \\
&\quad + \sum_{j=1}^{J} \left\lbrace (M_0\mu_0 -1)E_q  \left[ \log \mu_j \right] + (M_0 ( 1 - \mu_0) - 1) E_q  \left[ \log (1 - \mu_j)\right]\right\rbrace.
\end{split}
\end{equation}
Finally, since $r_{ji}|n_{ji}\thicksim \text{Binomial}(\theta_{ji}, n _{ji})$,
\begin{equation}
\begin{split}
E_q \left[ \log p\left(r | \theta, n \right)\right] &= \sum_{j=1}^{J} \sum_{i=1}^{N} E_q  \left[ \log p \left( r_{ji} | \theta_{ji}, n_{ji} \right) \right] \\
&= \sum_{j=1}^{J} \sum_{i=1}^{N}  E_q  \left[ \log \left( \frac{ \Gamma(n_{ji}+1) } { \Gamma(r_{ji}+1) \Gamma( n_{ji} - r_{ji} + 1 ) } \theta_{ji}^{r_{ji}} (1 - \theta_{ji})^{n_{ji} - r_{ji}} \right) \right] \\
&= \sum_{j=1}^{J} \sum_{i=1}^{N} \log \left( \frac{ \Gamma(n_{ji}+1) } { \Gamma(r_{ji}+1) \Gamma( n_{ji} - r_{ji} + 1 ) }\right)  \\
&\quad + \sum_{j=1}^{J} \sum_{i=1}^{N}  E_q  \left[ r_{ji} \log \theta_{ji} + (n_{ji} - r_{ji}) \log (1 - \theta_{ji}) \right] \\
&= \sum_{j=1}^{J} \sum_{i=1}^{N} \log \left( \frac{ \Gamma(n_{ji}+1) } { \Gamma(r_{ji}+1) \Gamma( n_{ji} - r_{ji} + 1 ) }\right)  \\
&\quad + \sum_{j=1}^{J} \sum_{i=1}^{N} \left\lbrace r_{ji} E_q \left[ \log \theta_{ji} \right] + (n_{ji} - r_{ji}) E_q  \left[  \log (1 - \theta_{ji}) \right] \right\rbrace. \\
\end{split}
\end{equation}

\bibliographystyle{named}
\bibliography{ref}

\end{document}